\documentclass[utf8]{frontiersSCNS} 

\usepackage{url,lineno,microtype,subcaption}
\usepackage[onehalfspacing]{setspace}
\usepackage{graphicx}
\usepackage{mathrsfs}
\usepackage{amsmath}
\usepackage{natbib}
\usepackage{color}
\usepackage{changes}

\def\keyFont{\fontsize{8}{11}\helveticabold }
\def\firstAuthorLast{Pian} 
\def\Authors{Elena Pian$^{1,*}$}
\def\Address{$^{1}$INAF, Astrophysics and Space Science Observatory}
\def\corrAuthor{Via Piero Gobetti 101, 40129 Bologna, Italy}
\def\corrEmail{elena.pian@inaf.it}

\begin{document}
\onecolumn
\firstpage{1}

\title[Binary neutron star mergers]{Mergers of binary neutron star systems: a multi-messenger revolution} 

\author[\firstAuthorLast ]{\Authors} 
\address{} 
\correspondance{} 

\extraAuth{}

\maketitle

\begin{abstract}

On 17 August 2017, less than two years after the direct detection of gravitational radiation from the merger of two $\sim$30 M$_\odot$ black holes, a binary neutron star merger was identified as the source of a gravitational wave signal of $\sim$100 s duration that occurred at less than 50 Mpc from Earth.  A short GRB was independently identified in the same sky area  by the  {\it Fermi} and {\it INTEGRAL}  satellites for high energy astrophysics, which turned out to be associated with the gravitational event.  Prompt follow-up observations at all wavelengths led first to the detection of an optical and infrared source located in the spheroidal galaxy NGC4993 and, with a delay of  $\sim$10 days,  to the detection of radio and X-ray signals.  This paper revisits these observations and focusses on the early optical/infrared source, which was thermal in nature  and powered by the radioactive decay of the unstable  isotopes of elements  synthesized via rapid neutron capture during the merger and in the phases immediately following it.  The far-reaching consequences of this event for cosmic nucleosynthesis and for the  history of heavy elements formation in the Universe are also illustrated.
\tiny
 \keyFont{ \section{Keywords:} neutron star, gravitational waves, gamma-ray burst, nucleosynthesis, r-process, kilonova} 
\end{abstract}


\section{Introduction}\label{sec:intro}

Although the birth of "multi-messenger" astronomy dates back to the detection of the first solar neutrinos in the '60s and was rejuvenated by the report of MeV neutrinos from SN1987A in the Large Magellanic Cloud,   the detection of gravitational radiation from the binary neutron star merger on 17 August 2017 (GW170817A) marks the transition to maturity of this approach to observational astrophysics, as it is expected to open an effective  window into the study of astrophysical sources  that is not limited to exceptionally close (the Sun) or rare (Galactic supernova) events. GW170817 is  a textbook case for gravitational physics, because, with its accompanying short gamma-ray burst (GRB) and afterglow, and its  thermal aftermath ("kilonova"), it has  epitomized the different epiphanies of the coalescence of a binary system of neutron stars, and finally allowed us to unify them.

Owing its name to a typical peak luminosity of $\sim 10^{42}$ erg~s$^{-1}$, i.e.  1000 times larger  than that of a typical nova outburst,  kilonova is the characteristic  optical and infrared source  accompanying a binary neutron star merger and due to the  radioactive decay of the many unstable isotopes of  large atomic weight elements  synthesized  via rapid neutron capture in the promptly formed dynamical ejecta and in the delayed post-merger ejecta.   Its evolution, as well as that of the GRB afterglow, was recorded with exquisite detail,  thanks to its closeness (40 Mpc).    Scope of this paper is to review the electromagnetic multi-wavelength observations of GW170817 with particular attention to the kilonova phenomenon.

The outline of the paper is as follows:  
Section \ref{sec:bnsm} sets the context of binary systems of neutrons stars and describes the predicted  outcomes of their coalescences;
Section \ref{sec:gw170817} presents the case of GW170817, the only so far confirmed example of double neutron star merger and the multi-wavelength features of its electromagnetic counterpart (short GRB and kilonova); 
Section \ref{sec:rddcyhel}  focusses on the kilonova, elaborates on its observed optical and near-infrared light curves and spectra, draws the link  with nucleosynthesis of heavy elements, and outlines  the  theoretical framework that is necessary to describe the kilonova properties and implications;  
Section \ref{sec:concl}   summarizes the results and provides an outlook of this line of research in the near future.


\section{Binary neutron star mergers}\label{sec:bnsm}

Neutron stars are  the endpoints of massive stars evolution and therefore  ubiquitous in the Universe:
on average, they represent about 0.1\% of the total stellar content of a galaxy. Since massive stars are mostly in binary systems \citep{Sana2012}, neutron star binaries should form  readily,  if the supernova explosion of either progenitor massive star does not disrupt the system \citep{Renzo2019}.    Alternatively, binary neutron star systems can form dynamically  in dense environments like stellar clusters (see Ye et al., 2020, and references therein).   Binary systems composed by a neutron star and a black hole are also viable, but rare \citep{Pfahl2005}, which may account for the fact that none has so far been detected in our Galaxy.

The prototype binary neutron star system in our Galaxy is  PSR B1913+16, where one member was detected as a  pulsar  in a  radio survey carried out at the Arecibo  observatory \citep{HulseTaylor1974}, and the presence of its companion was inferred from the periodic changes in the observed pulsation period of 59 ms \citep{HulseTaylor1975}.   Among various tests of strong general relativity enabled by the radio monitoring of this binary system, which earned the Nobel Prize for Physics to the discoverers in 1993, was the measurement of the shrinking of the binary system orbit, signalled by the secular decrease  of the 7.75 hours orbital period,  that could be entirely attributed to energy loss via gravitational radiation (Taylor \& Weisberg, 1982; Weisberg \& Huang,  2016, and references therein).

With an orbital decay rate of $\dot{P} = -2.4 \times 10^{-12}$ s s$^{-1}$, the merging time of the PSR B1913+16 system is $\sim$300 Myr.   Following the detection of PSR B1913+16,  another dozen  binary neutron stars systems were detected in our Galaxy  (e.g. Wolszczan, 1991; Burgay et al.,  2003;  Tauris et al., 2017; Martinez et al., 2017).   Almost half  of these have estimated merging times significantly shorter than a Hubble time.    
The campaigns conducted by the LIGO  interferometers  in Sep 2015-Jan 2016 (first observing  run) and, together with Virgo, in Nov 2016-Aug 2017 (second observing run), the latter  leading to the  first detection of gravitational waves from a merging double neutron star system (see Section \ref{sec:gw170817}),  constrained the local merger rate density to be   110-3840  Gpc$^{-3}$ yr$^{-1}$  \citep{Abbottapj2019}. This  is consistent with previous estimates (see e.g. 
Burgay et al.,  2003), and, under a series of assumptions,  marginally consistent with  independent estimates based on  double neutron star system formation in the classical binary evolution scenario \citep{Chruslinska2018}.   Ye et al. (2020) have estimated that the fraction of merging binary neutron stars that have formed dynamically in globular clusters is negligible.   Under the assumption that the event detected by LIGO on 25 April 2019  was produced by a binary neutron star coalescence  the local rate of neutron star mergers would be updated to $250-2810$ Gpc$^{-3}$ yr$^{-1}$  \citep{Abbottapj2020a}.
 
The merger of a binary neutron star system has four predicted outcomes: 1) a gravitational wave signal that is mildly isotropic, with a stronger intensity in the polar direction  than in the equatorial plane;  2) a  relativistic outflow, which is highly anisotropic and can produce an observable high energy transient;   3) a  thermal, radioactive source emitting most of its energy  at  ultraviolet, optical and near-infrared wavelengths; 4) a burst of MeV neutrinos \citep{Eichler1989,
RosswogLiebendoerfer2003} following the formation of the central remnant, and possibly of high-energy ($>$GeV) neutrinos from hadronic interactions within the relativistic jet \citep{FangMetzger2017,Kimura2018}.  While neutrinos are  extremely elusive and detectable only from very small distances with present instrumentation (see Section \ref{sec:concl}), the first three observables have been now all detected, as detailed in the  next three sub-sections.

\subsection{Gravitational waves}\label{sec:gws}

Coalescing binary systems of degenerate stars and stellar mass black holes are optimal candidates for the generation of  gravitational waves detectable from ground-based interferometers  as the strong gravity conditions lead to huge velocities and energy losses (Shapiro \& Teukolsky, 1983), and the frequency of the emitted gravitational waves  reaches several kHz,  where the sensitivity of the advanced LIGO, Virgo and KAGRA  interferometers is designed to be maximal \citep{Abbott2018}.    

The time behavior of binary systems  of compact stars consists of three phases: a first inspiral phase in a close orbit  that shrinks as gravitational radiation of frequency proportional to the orbital frequency is emitted, a merger phase where a remnant compact body is produced as a result of the coalescence of the two stars, and a post-merger, or ringdown, phase where the remnant  still emits gravitational radiation while settling to its new stable configuration.    During the inspiral, the amplitude of the sinusoidal gravitational signal rapidly increases as the distance between the two bodies decreases and the frequency increases (chirp), while in the ringdown phase  the signal is an exponentially damped sinusoid.  This final phase may encode critical information on the equation of state of the newly formed remnant (a black hole or, in the case of light neutron stars, a massive neutron star or metastable supramassive neutron star).   The  mathematical tool that is used to describe this evolution is the waveform model, that aims at reproducing the dynamics of the system through the application of post-Newtonian corrections of increasing order and at providing the essential parameters that can then  be compared with the interferometric observations \citep{Blanchet2014,Nakano2019}.   

Since the amplitude of gravitational waves  depends on the masses of the binary member stars, the signal will be louder, and thus detectable from larger distances, for binary systems that involve black holes than those with neutron stars.   The current  horizon for binary neutron star merger detection with LIGO  is $\sim$200 Mpc, and 25-30\% smaller with Virgo and  KAGRA (Abbott et al. 2018).   The dependence of the gravitational waves amplitude on the physical parameters of the system implies that gravitational wave sources are standard sirens \citep{Schutz1986}, provided account is taken of the correlation between the luminosity distance and the inclination of the orbital plane with respect to the line of sight \citep{Nissanke2010,Abbott2016}.

\subsection{Short gamma-ray bursts}\label{sec:sgrbs}

GRBs,  flashes of radiation of 100-1000 keV that  outshine the entire Universe in this band, have durations between a fraction of a second and hundreds or even thousands of seconds.  However, the duration distribution is bimodal, with a peak around 0.2 seconds (short or sub-second GRBs) and one around 20 seconds (long GRBs; Kouveliotou et al., 1993).   This bimodality is reflected in the spectral hardness, which is on average larger in short GRBs,  and in a  physical difference between the two groups.   While most long GRBs are associated with core-collapse supernovae \citep{Galama1998,WoosleyBloom2006,Levan2016},  sub-second GRBs are produced by the merger of two neutron stars or a neutron star and a black hole, as long predicted based on circumstantial evidence \citep{Blinnikov1984,Eichler1989,Fong2010,Berger2013,Tanvir2013}  and then proven by the detection  of GW170817  and of its high energy counterpart GRB170817A  (Section \ref{sec:gw170817}). 
The observed relative ratio of long versus short GRBs depends on the detector sensitivity and effective energy band (e.g. Burns et al., 2016).  However,  the duration overlap of the two populations is very large, so that the minimum of the distribution  has to be regarded as a rather vaguely defined value \citep{Bromberg2013}.  

About 140 short GRBs were localized so far to a precision that is better than 10 arc-minutes\footnote{http://www.mpe.mpg.de/$\sim$jcg/grbgen.html};  of these,  $\sim$100, $\sim$40 and $\sim$10 have a detected afterglow in X-rays, optical and radio wavelengths, respectively, and $\sim$30 have measured redshifts (these range between $z = 0.111$ and $z = 2.211$, excluding the nearby GRB170817A, see Section \ref{sec:grb170817a}, and GRB090426, $z = 2.61$, whose identification as a short GRB is not robust, Antonelli et al., 2009).    Short GRBs are located at projected  distances of a fraction of, to several kiloparsecs from, the centers of their host galaxies, which are of both early and late type, reflecting the long time delay between the formation of the short GRB progenitor binary systems and their mergers \citep{Berger2014}.

According to the classical  fireball model, both prompt event and multi-wavelength afterglow  of short GRBs are produced in a highly relativistic jet directed at a small angle with respect to the line of sight, whose aperture can be derived from the achromatic steepening (or "jet break") of the observed afterglow light curve \citep{Nakar2007}.    In principle, this could be used to reconstruct the collimation-corrected rate of short GRBs, to be compared with predictions of binary neutron star merger rates.  However, these estimates proved to be very uncertain, owing to the difficulty of measuring accurately the jet breaks in short GRB afterglows \citep{Fong2015,Jin2018,Wang2018,Lamb2019,Pandey2019}.

\subsection{r-process nucleosynthesis}\label{sec:rprocess}

Elements heavier than iron cannot form via stellar nucleosynthesis, as not enough neutrons are available for the formation of nuclei and temperatures are not sufficiently high to overcome the repulsive Coulomb barrier that prevents acquisition of further baryons into nuclei \citep{Burbidge1957}.    Supernovae (especially the thermonuclear ones) produce large amounts of iron via decay (through  $^{56}$Co) of radioactive $^{56}$Ni synthesized  in the explosion.   Heavier nuclei form  via four  neutron capture processes \citep{Thielemann2011},  the dominant ones being slow and rapid neutron capture, in brief s- and r-process, respectively, where "slow" and "rapid" refer to the timescale  of neutron accretion into the nucleus with respect to that of the competing process of $\beta^-$ decay.   In the s-process, neutron captures occur with timescales of hundreds to thousands of years, making $\beta^-$ decay highly probable, while r-process neutron capture occurs on a  timescale of $\sim$0.01 seconds, leading to acquisition of many neutrons before $\beta^-$ decay can set on.  As a consequence, the s-process produces less unstable, longer-lived isotopes, close to the so-called  valley of  $\beta$-stability (the decay time of a radioactive nucleus  correlates inversely with its number of neutrons), while the r-process produces the heaviest, neutron-richest and most unstable isotopes of heavy nuclei, up to uranium  \citep{Sneden2008,MennekensVanbeveren2014,Thielemann2017,Cote2018,Horowitz2019,Kajino2019,Cowan2020}.   Among both s-process and r-process elements, some are particularly stable owing to their larger binding energies per nucleon, which causes their abundances to be relatively higher than others.  In the abundances distribution in the solar neighbourhood these are seen as maxima ("peaks")  centered around atomic numbers $Z$ = 39 (Sr-Y-Zr) , 57 (Ba-La-Ce-Nd), 82 (Pb)  for the s-process and, correspondingly somewhat lower atomic numbers $Z$ = 35 (Se-Br-Kr), 53 (Te-I-Xe), 78 (Ir-Pt-Au)  for  the r-process  (e.g. Cowan et al., 2020).

Both s-process and r-process naturally occur in environments that are adequately supplied with large neutron fluxes.  For the s-process, these are eminently asymptotic giant branch stars, where neutron captures are driven by the $^{13}$C($\alpha$, n)$^{16}$O and  $^{22}$Ne($\alpha$, n)$^{25}$Mg   reactions   \citep{Busso1999}.  The r-process requires much higher energy and neutron densities,  which are only realized in  most physically extreme environments.  While it can be excluded that big-bang nucleosynthesis can accommodate  heavy elements formation in any significant amount  \citep{Rauscher1994}, there is currently no consensus on the relative amounts of  nucleosynthetic yields in the prime  r-process  candidate   sites:  core-collapse supernovae and mergers  of binary systems composed by neutron stars or a  neutron star and a black hole.

Core-collapse supernovae have been proposed starting many decades ago as sites of r-process nucleosynthesis through various mechanisms  and in different parts of the explosion, including dynamical ejecta of prompt explosions of O-Ne-Mg cores \citep{Hillebrandt1976,Wheeler1998,Wanajo2002};  C+O layer of O-Ne-Mg-core  supernovae \citep{Ning2007}; He-shell exposed to intense neutrino flux \citep{Epstein1988,Banerjee2011}; re-ejection of fallback material  \citep{Fryer2006}; neutrino-driven wind from proto-neutron stars  \citep{Woosley1994,Takahashi1994};  magnetohydrodynamic jets of rare core-collapse SNe  \citep{Nishimura2006,Winteler2012}.  Similarly old is the first proposal that  the tidal disruption of neutron stars by black holes in close binaries \citep{Lattimer1974,Lattimer1976,SymbalistySchramm1982,Davies1994}  and  coalescences of binary neutron star systems  \citep{Eichler1989} could be at the origin of r-process nucleosynthesis.   This should manifest as a thermal optical-infrared source of radioactive nature of much lower luminosity   (a factor of 1000)   and shorter duration  (rise time of a  few days) than supernova \citep{LiPaczynski1998}.   

The models for r-process elements production in core-collapse supernova  all have problems inherent their physics (mostly  related to energy budget and neutron flux density).  On the other hand, the binary compact star merger origin may fail to explain observed r-process element abundances in very low metallicities stars, i.e. at very early cosmological epochs, owing to the non-negligible binary evolution times (see Cowan et al., 2020 for an accurate review of all arguments in favour and against either channel).  While  the event of 17 August 2017 (Section \ref{sec:gw170817}) has now provided incontrovertible evidence that binary neutron star mergers host r-process nucleosynthesis, the role of core-collapse supernovae cannot be dismissed although their relative contribution with respect to the binary compact star channel must be assessed \citep{RamirezRuiz2015,Ji2016,Shibagaki2016,Cote2019,Safarzadeh2019,Simonetti2019}.  It cannot be excluded that both  "weak" and  "strong" r-process nucleosynthesis takes place, with the former occurring mainly in supernova and possibly failing to produce atoms up to the third peak of r-process elemental abundance distribution \citep{Cowan2020}.    The hint that  heavy elements may be produced in low-rate events with high yields \citep{Sneden2008,Wallner2015,MaciasRamirezRuiz2019} points to binary compact star mergers or very energetic (i.e.  expansion velocities larger than 20000 km s$^{-1}$)  core-collapse supernovae  as progenitors, rather than regular core-collapse supernovae.  Along these lines, it has been proposed that  accretion disks of collapsars (the powerful core-collapse supernovae that accompany long GRBs, Woosley \& Bloom, 2006)  produce
neutron-rich outflows that synthesize heavy r-process nuclei \citep{Nakamura2013,Kajino2014,Nakamura2015}.
\citet{Siegel2019}  calculated that collapsars may supply more than 80\%  of the r-process content and computed  synthetic spectra for models of r-process-enriched  supernovae corresponding to an MHD supernova and a collapsar
disk outflow scenario.

Neutrons are tightly packed together in neutrons stars, but  during coalescence of a binary neutron star system the 
tidal forces disrupt them and the released material forms promptly a disk-like rotating structure (dynamical ejecta, Rosswog et al., 1999; Shibata \& Hotokezaka, 2019) where the neutron density rapidly drops to optimal values for  r-process occurrence ($\sim 10^{24-32}$ neutrons~cm$^{-3}$, Freiburghaus et al., 1999) and  for copious formation of neutron-rich stable and unstable isotopes of large atomic number elements \citep{FernandezMetzger2016,Tanaka2016,Tanaka2018,Wollaeger2018,Metzger2019}.


\section{The binary neutron star merger of 17 August 2017}
\label{sec:gw170817}

On 17 August 2017, the LIGO and Virgo interferometers detected for the first time a gravitational signal that corresponds to the final inspiral and coalescence of a binary neutron star system \citep{Abbottprl2017a}.   The sky uncertainty area associated with the event was 28 square degrees, in principle too large for a uniform search for an  electromagnetic counterpart with ground-based and orbiting telescopes.  However, its small distance  ($40^{+8}_{-14}$  Mpc), estimated via  the "standard siren"  property of gravitational wave signals associated with binary neutron star mergers,  suggested that the aftermath could be rather bright, and motivated a  large-scale campaign at all wavelengths from radio to very high energy gamma-rays,  which was promptly and largely rewarded by success and then timely followed by a long and intensive monitoring \citep{Abbottapj2017b,Abbottapj2017c}, as described in Section \ref{sec:gwemcp}.   Searches of MeV-to-EeV neutrinos directionally coincident with the source using data from the Super-Kamiokande, ANTARES, IceCube, and Pierre Auger Observatories between 500 seconds before and 14 days after the merger returned no detections \citep{Albert2017,Abe2018}.

Based on the very detection of electromagnetic radiation, \cite{Bauswein2017}  have argued that the merger remnant may not be a black hole or at least the post-merger collapse to a black hole may  be delayed.    Since the post-merger phase ("ring-down") signal of GW170817 was not detected \citep{Abbottapj2017e}, this hypothesis cannot be tested directly with gravitational data.  \cite{Bauswein2017} also derived lower limits on the radii of the neutron stars.  

Notably, while the gravitational data made it possible to set an upper limit on the tidal-deformability  parameter of the binary neutron stars ($\tilde\Lambda \lesssim 800$, Abbott et al. 2017a), the optical  observation of kilonova ejecta limited the same parameter from below ($\tilde\Lambda \gtrsim 400$, Radice et al. 2018), based on the consideration that for smaller values of $\tilde\Lambda$ a long-lived remnant would not be favoured, contradicting the result of \citet{Bauswein2017}.   The limits on the $\tilde\Lambda$ parameter constrain  the neutron star radius to the range 11.8 km $\lesssim R_{1.5} \lesssim 13.1$ km, where $R_{1.5}$ refers to a 1.5 $M_{\odot}$ neutron star \citep{Burgio2018}, and 
in turn confine the possible ensemble of viable equations of state \citep{Annala2018,LimHolt2018}, a fundamental, yet poorly known descriptor of neutron star physics \citep{OzelFreire2016}.    Furthermore, by circumscribing the number of equations of state of the compact stars, their exploration can be brought beyond nucleonic matter,  and extended to scenarios of  matter presenting  a phase transition \citep{Burgio2018,Most2018}.
The results on the tidal-deformability of the neutron star progenitors of GW170817 and on the behavior of the remnant thus provide a brilliant confirmation of the added value of a multi-messenger approach over separate observations of individual carriers of information.

\subsection{The electromagnetic counterpart of GW170817}\label{sec:gwemcp}

Independent of  LIGO-Virgo detection of the gravitational wave signal, the Gamma-ray Burst Monitor (GBM) onboard the  NASA {\it Fermi} satellite  and the  Anticoincidence Shield for the gamma-ray Spectrometer  (SPI) of the {\it International Gamma-Ray Astrophysics Laboratory} ({\it INTEGRAL}) satellite  were  triggered by a faint short GRB (duration of $\sim$2 seconds),  named GRB170817A  \citep{Abbottapj2017b,Goldstein2017,Savchenko2017}.   This gamma-ray transient, whose large error box was compatible with that determined by LIGO-Virgo, lags the gravitational merger by 1.7 seconds, a delay that may be dominated by the propagation time of the jet to the gamma-ray production site (Beniamini et al., 2020; see however Salafia et al., 2018).
The preliminary estimate of the  source distance provided a crucial constraint on the maximum distance of the galaxy that could plausibly have hosted the merger, so that the searching strategy was based on targeting galaxies within a $\sim$50 Mpc cosmic volume (see e.g. Gehrels et al., 2016) with telescopes equipped with large (i.e. several square degrees) field-of-view cameras.   

About 70 ground-based optical telescopes participated to the hunt and each of them adopted a different pointing sequence.    This systematic approach enabled  many groups to identify the optical counterpart candidate in a timely manner (with optical magnitude $V \simeq 17$), i.e. within $\sim$12 hours of the merger  \citep{Arcavi2017,Lipunov2017,SoaresSantos2017,Valenti2017,Tominaga2018}.  \cite{Coulter2017}  were the first to report a detection with the optical 1m telescope Swope at Las Campanas Observatory.  The optical source lies at 10 arc-seconds  angular separation, corresponding to a projected   distance of $\sim$2 kpc, from the center of  the spheroidal galaxy NGC~4993 at 40 Mpc  \citep{Blanchard2017,Im2017,Levan2017,Pan2017,Tanvir2017}. 

Rapid follow-up of the gravitational wave and GRB signal in X-rays did not show any source comparable to, or brighter than a typical afterglow of a short GRB.   Since both the gravitational data and the faintness of the prompt GRB emission suggested a jet viewed significantly off-axis, this  could be expected,  as the afterglows from misaligned GRB jets have longer rise-times  than those of jets observed at small viewing angles  \citep{VanEertenMacFadyen2011}.
Therefore, X-ray monitoring with {\it Swift}/XRT,  {\it Chandra}  and {\it Nustar} continued,  and $\sim$10 days after merger led to the detection  with {\it Chandra}    of a faint source ($L_X \simeq 10^{40}$ erg~s$^{-1}$) \citep{Evans2017,Margutti2017,Troja2017}, whose intensity continued to rise up to $\sim$100 days  \citep{Davanzo2018,Troja2020}. 
Similarly, observations at cm and mm wavelengths at various arrays, including VLA and ALMA, failed to detect the source before $\sim$16 days after the gravitational signal, which was interpreted as evidence that a jetted source accompanying the binary neutron star merger must be directed at a significant angle ($\ge$20 degrees) with respect to the line of sight \citep{Alexander2017,Andreoni2017,Hallinan2017,Kim2017,Pozanenko2018}.

The  {\it Fermi}  Large Area Telescope covered the sky region of GW170817 starting only 20 minutes after the merger, and did not detect any emission in the energy range 0.1-1 GeV  to a limiting flux of 
$4.5 \times 10^{-10}$  erg~s$^{-1}$~cm$^{-2}$ in the interval 1153-2027 seconds after the merger \citep{Ajello2018}.  Follow-up observation with the atmospheric Cherenkov experiment H.E.S.S. (0.3-8 TeV) from a few hours to $\sim$5 days after merger returned no detection to a limit of a few 10$^{-12}$ erg~s$^{-1}$~cm$^{-2}$ \citep{Abdalla2017}.    A  summary of the results of the multi-wavelength observing  campaign within the first month of gravitational wave signal detection  are reported in  \cite{Abbottapj2017c}.

While the radio and X-ray detections are attributed to the afterglow of the short GRB, the ultraviolet, optical and near-infrared data are dominated  by the kilonova at early epochs (with a possible contribution at $\lesssim$4 days at blue wavelengths from cooling of shock-heated material around the neutron star merger, Piro \& Kollmeier, 2018),  and later on by the afterglow, as described in the next two Sections.

\subsubsection{The  gamma-ray burst and its multi-wavelength  afterglow}\label{sec:sgrbag}
\label{sec:grb170817a}

The short  GRB170817A, with an energy output of  $\sim 10^{46}$ erg,  was orders of magnitude dimmer than most short GRBs \citep{Berger2014}.   Together with a viewing angle of $\sim$30 deg estimated from the gravitational wave signal \citep{Abbottprl2017a}, this led to the hypothesis that the GRB was produced by a relativistic jet viewed at a comparable angle.   However,  the early light curve of  the radio afterglow  is not consistent with the behavior predicted for an off-axis collimated jet and rather suggests a quasi-spherical geometry, possibly with two components, a more collimated one and a nearly isotropic and mildly relativistic one, which is responsible also for producing the gamma-rays   \citep{Mooley2018a}.    This confirms numerous predictions whereby the shocked cloud surrounding 
a binary neutron star merger forms a mildly relativistic cocoon  that carries an energy comparable to that of the jet and is responsible for the prompt emission and the early multi-wavelength afterglow  \citep{Lazzati2017a,Lazzati2017b,NakarPiran2017,Bromberg2018,Xie2018},  and is supported by detailed numerical simulations \citep{Lazzati2018,Gottlieb2018}.   

Using milli-arcsecond resolution radio VLBI observations  at 75 and 230 days Mooley et al. (2018b)  detected superluminal motion with $\beta = 3-5$,  while Ghirlanda et al. (2019) determine that, at 207 days, the source is still angularly smaller than 2 milli-arcseconds at the
90\% confidence,  which  excludes  that a nearly isotropic, mildly relativistic outflow is responsible for
the radio emission, as in this case  the source apparent size, after more than six months of expansion, should  be significantly larger and resolved by the VLBI observation.   These observations point to a structured jet as the source of GRB170817A, with a narrow opening angle ($\theta_{op} \simeq 3.4$ degrees)  and an energetic core ($\sim 3 \times 10^{52}$ erg) seen under a viewing angle of $\sim$15 degrees \citep{Ghirlanda2019}.   This is  further  confirmed  by later radio observations extending up to 300 days after merger, that show a sharp downturn of the radio light curve, suggestive of a jet rather than a spherical source \citep{Mooley2018c}.

The optical/near-infrared  kilonova component subsided rapidly (see Section \ref{sec:kn})  leaving room to the afterglow emission: the  HST observations at $\sim$100 days after the explosion show a much brighter source than inferred from the extrapolation of the early kilonova curve to that epoch \citep{Lyman2018}.  This late-epoch flux is thus not consistent with kilonova emission and rather due to the afterglow produced within an off-axis structured  jet \citep{Fong2019}.
At X-ray energies, the GRB counterpart is still detected with {\it Chandra} three  years after explosion \citep{Troja2020}, but its decay is not fully compatible  with a structured jet, indicating that the physical conditions have changed or that an extra component is possibly emerging (e.g. a non-thermal aftermath of the  kilonova ejecta, see next Section).

\subsubsection{The  kilonova}\label{sec:kn}

The early ground-based optical and near-infrared  and space-based (with {\it Swift}/UVOT) near-ultraviolet follow-up observations  started immediately after identification of the optical counterpart of GW170817, detected a rapid rise ($\sim$1 day timescale, Arcavi et al. 2017) and wavelength-dependent time decay, quicker at shorter wavelengths   \citep{Andreoni2017,Cowperthwaite2017,Diaz2017,Drout2017,Evans2017, McCully2017,Nicholl2017,Tanvir2017,Utsumi2017,Villar2017}.    
The optical light is  polarized  at the very low  level of ($0.50 \pm 0.07$)\% at 1.46 days, consistent with intrinsically unpolarized emission scattered by Galactic dust,  indicating that no significant alignment effect in the emission or geometric preferential direction is present in the source at this epoch, consistent with expectation for kilonova emission \citep{Covino2017}.

Starting the same night when the optical counterpart was detected,  low resolution spectroscopy was carried out at the 
Magellan telescope \citep{Shappee2017}.   This spectrum shows that the source is not yet transparent as it is emitting  black body radiation, whose maximum lies however blue-ward of the sampled wavelength range,  suggesting that the initial temperature may have been larger than $\sim$10000~K.  The following night (1.5 days after merger)  the spectrum is still described by an almost perfect black body law whose maximum at $\sim$5000 K was fully resolved by  spectroscopy at the Very Large Telescope (VLT) with the X-Shooter spectrograph over the wavelength range 3500-24000 \AA\  \citep{Pian2017}.     At this epoch, the expansion velocity of the expelled ejecta, whose total mass was  estimated  to be   0.02-0.05 M$_\odot$  \citep{Pian2017,Smartt2017,Waxman2018},  was   $\sim$20\% of the light speed, 
which is only mildly relativistic and therefore much less extreme than the ultra-relativistic kinematic regime of the GRB and of its early afterglow,  analogous to the observed difference between the afterglows and the supernovae accompanying long GRBs.   At 2.5 days after merger the spectrum starts deviating from a black body as the ejecta become increasingly   transparent and absorption lines are being  imprinted on the spectral continuum by the atomic species present in the ejecta \citep{Chornock2017,Pian2017,Smartt2017}.   In the following days these  features  become prominent and they evolve as the ejecta decelerate and the photosphere recedes (Figure \ref{fig:1}).   

In particular, in the spectrum at day 1.5 an absorption feature extending from  $\sim$7000 \AA\  to  $\sim$8100 \AA\  is detected, that \citet{Smartt2017}  preliminarily identified with atomic transitions occurring in neutral Cs and Te, broadened and blueshifted by  $\sim 0.2c$,  consistent with the expansion velocity of the photosphere.  In the second spectrum (2.5 days) the Cs {\sc I} and Te {\sc I}  lines are still detected at somewhat larger wavelengths, compatibly with a reduced photospheric expansion speed.  These lines were however  later  disproved  on account of the fact that,  at the temperature of the ejecta immediately below the photosphere ($\sim$3700 K), numerous transitions of other lanthanide elements of higher ionisation potential should be observed besides Cs and Te, but are not \citep{Watson2019}.   \citet{Watson2019} reanalysed the absorption feature observed at 7000-8100 \AA\ and an absorption feature at $\sim$3500 \AA\   with the aid of local thermodynamic equilibrium  models with abundances from a solar-scaled r-process and from metal-poor stars and determined that the absorption features can be identified with Sr {\sc II}.   In the spectra at the successive epochs the line at the longer wavelength is still detected and  develops  a P Cygni profile. Strontium is a very abundant element and is produced close to the first r-process peak.  Its possible detection makes it important to consider lighter r-process elements in addition to the lanthanides in shaping the kilonova emission spectrum \citep{Watson2019}.

At  $\sim$10 days after merger, the kilonova spectrum fades out of the reach of the largest telescopes.   The radioactive source could still be monitored photometrically for another week in optical and near-infrared \citep{Cowperthwaite2017,Drout2017,Kasliwal2017,Pian2017,Smartt2017,Tanvir2017};  it was last detected at 4.5 $\mu$m with the  {\it Spitzer}  satellite  74 days post merger \citep{Villar2018}.
The kilonova ejecta are also expected to interact with the circum-binary medium  and produce low-level radio and X-ray emission that peaks years after the merger \citep{Kathirgamaraju2019}.   Search for this component has not returned (yet) a detection at radio wavelengths \citep{Hajela2019}, but it may start to be revealed at X-rays \citep{Troja2020}.

\subsubsection{The host galaxy of GW170817}\label{sec:hostgal}

HST and {\it Chandra} images, combined with VLT MUSE integral field spectroscopy of the optical counterpart of  GW170817, show that its host galaxy, NGC 4993, is a lenticular (S0)  galaxy at $z = 0.009783$ that has undergone a recent ($\sim$1 Gyr) galactic merger \citep{Levan2017,Palmese2017}.   This merger may be responsible for igniting  weak nuclear activity.    No globular or young stellar cluster is detected at the location of GW170817, with a limit of a few thousand solar masses for any young system. The population in the vicinity is predominantly old and the extinction from local interstellar medium low.     Based on these data, the distance of NGC4993 was determined to be  ($41.0 \pm 3.1$) Mpc \citep{Hjorth2017}.   The HST imaging  made it also possible to establish  the distance of NGC4993  through the surface brightness fluctuation method  with an uncertainty  of $\sim$6\% ($40.7 \pm 1.4 \pm 1.9$ Mpc, random and systematic errors, respectively),  making it the most precise distance measurement for this galaxy \citep{Cantiello2018}.  
Combining this with the  recession velocity measured from optical spectroscopy of the galaxy, corrected for peculiar motions,  returns a Hubble constant $H_0 = 71.9 \pm 7.1$ km~s$^{-1}$~Mpc$^{-1}$. 

Based only on the gravitational data and the standard siren argument, and assuming that the optical counterpart represents
the true sky location of the gravitational-wave source instead of marginalizing over a range of potential sky locations, \cite{Abbottnat2017d} determined a "gravitational" distance of $43.8^{+2.9}_{-6.9}$ Mpc that is refined with respect to the one previously  reported in \cite{Abbottprl2017a}.  Together with the corrected recession velocity of NGC4993 this yields a  Hubble constant $H_0 = 70^{+12}_{-8}$ km~s$^{-1}$~Mpc$^{-1}$, comparable to, but less precise than that obtained from the superluminal motion of the radio counterpart core,  $H_0 = 70.3^{+5.3}_{-5.0}$  km~s$^{-1}$~Mpc$^{-1}$ \citep{Hotokezaka2019}.


\section{Kilonova light curve and spectrum}\label{sec:rddcyhel}
  
The unstable isotopes formed during coalescence of a binary neutron star system decay radioactively and the emitted gamma-ray photons are down-scattered to the ultraviolet, optical and infrared thermal radiation that constitutes the kilonova  source  (Section \ref{sec:kn}).  Its time decline is determined by the convolution of  radioactive decay chain curves of all present unstable nuclei.   This is  analogous to the supernova phenomenon, where however the vastly dominant radioactive chain is $^{56}$Ni  decaying into  $^{56}$Co, and then into $^{56}$Fe.   

While radioactive nuclei decay, atoms recombine,  as the source is cooling,  and absorption features are imprinted in the kilonova spectra.  Among neutron-rich nuclei, the lanthanides (atomic numbers 57 to 71) series have full f-shells and therefore numerous atomic transitions that suppress  the spectrum at shorter wavelengths ($\lesssim$ 8000 \AA).   Spectra of dynamical ejecta of kilonova may therefore be heavily  intrinsically reddened, depending on the relative abundance of lanthanides \citep{BarnesKasen2013,Kasen2013,TanakaHotokezaka2013}.
Prior to the clear detection of kilonova accompanying GW170817 (Section \ref{sec:gw170817}),  such a source may have been detected in HST images in near-infrared H band of the afterglow of GRB130603B \citep{Berger2013,Tanvir2013}.   Successive claims for association with short GRBs and kilonova radiation were similarly uncertain \citep{Jin2015,Jin2016}.  

If the neutron stars coalescence does not produce instantaneously  a black hole, and  a  hypermassive neutron star is formed as a transitory remnant, a neutrino wind is emitted, that may inhibit the formation of neutrons and reduce the amount of neutron-rich elements \citep{FernandezMetzger2013,Kajino2014,Kiuchi2014,MetzgerFernandez2014,Perego2014,Kasen2015,Lippuner2017}.  This "post-merger" kilonova component, of preferentially polar direction, is thus relatively poor in lanthanides and gives rise to a less reddened spectrum \citep{Tanaka2017,Kasen2017}.

The optical/near-infrared spectral behavior of kilonova is analogous to that of supernovae with the largest kinetic energies  ($> 10^{52}$ erg), like those associated with GRBs: the large photospheric velocities  broaden the absorption lines and  blueshift them  in the direction of the observer.  Furthermore,  broadening causes the lines to blend, which  makes it difficult to isolate and identify individual atomic species \citep{Iwamoto1998,Mazzali2000,Nakamura2001}.  While these effects can be controlled and de-convolved  with the aid of a radiation transport model as it has been done  for supernovae of all types  \citep{Mazzali2016,Hoeflich2017,Ergon2018,HillierDessart2019,Shingles2020,AshallMazzali2020},  a more fundamental hurdle in modelling kilonova spectra consists in the much larger number of  electronic transitions  occurring in r-process element atoms than in the  lighter ones that populate supernova ejecta,  and in our extremely limited knowledge of individual atomic opacities of these neutron-rich elements, owing to the lack of suitable atomic data.
First systematic atomic structure calculations for lanthanides and for all  r-process elements were presented by  Fontes et al. (2020) and Tanaka et al. (2020), respectively.


\section{Summary  and future prospects}\label{sec:concl}

The gravitational and electromagnetic event of 17 August 2017 provided the  long-awaited confirmation that binary neutron star mergers are responsible for well identifiable gravitational signals at kHz frequencies, for short GRBs, and for thermal  sources, a.k.a. kilonovae or macronovae,  produced by the radioactive  decay of unstable heavy elements synthesized via r-process during the coalescence.   The intensive  and longterm electromagnetic monitoring from ground and space allowed clear detection of the  counterpart at all wavelengths.   Brief ($\sim$2 s) gamma-ray emission, peaking at $\sim$200 keV and lagging the gravitational signal by 1.7 seconds, is consistent with a weak short GRB.  At  ultraviolet-to-near-infrared wavelengths, the kilonova component - never before detected to this level of accuracy and robustness - dominates during the first 10 days and decays rapidly under detection threshold thereafter, while  an afterglow component emerges around day $\sim$100.    Up to the most recent epochs of observation (day $\sim$1000 at X-rays), the kilonova does not add significantly  to the bright radio and X-ray afterglow component.  Multi-epoch VLBI observations measured - for the first time in a GRB - superluminal motion of the radio source, thus providing evidence of late-epoch emergence of a collimated off-axis relativistic jet.   

Doubtlessly,  this series of breakthroughs were made possible  by the  closeness of the source (40 Mpc), almost unprecedented for GRBs, and by the availability of first-class  ground-based and space-borne  instruments.   The many findings and exceptional new physical insight afforded by GW170817/GRB170817A  make it a {\it rosetta stone} for future similar events.    When a sizeable group of  sources with good gravitational and electromagnetic  detections will be available, the  properties of binary  systems containing at least one neutron star, of their mergers and their aftermaths  can be mapped.  It will then become possible to clarify how the dynamically ejected mass depends on the binary system parameters, mass asymmetry and neutron stars equation of state \citep{RuffertJanka2001,Hotokezaka2013}, how the jet forms and evolves, which kinematic regimes and geometry it takes up in time, and how can the GRB and afterglow observed phenomenologies help distinguish the intrinsic properties from  viewing angle effects \citep{Janka2006,LambKobayashi2018,IokaNakamura2019}, what is the detailed chemical content of kilonova ejecta  and how the r-process abundance pattern inferred from kilonova spectra compares with the history of heavy elements cosmic enrichment \citep{Rosswog2018}, how can kilonovae help constrain the binary neutron star rates and how does the parent population of short GRBs evolve \citep{GuettaStella2009,Yang2017,Belczynski2018,Artale2019,MatsumotoPiran2020}, how gravitational and electromagnetic data can be used jointly to determine the cosmological parameters \citep{Schutz1986,DelPozzo2012,Abbottnat2017d},   to mention only some fundamental open problems.
Comparison of the optical and near-infrared  light curves of GW170817 kilonova with those of short GRBs with known redshift suggests infact significant diversity in the kilonova component luminosities \citep{Gompertz2018,Rossi2020}.

Regrettably,   short GRBs viewed at random angles, and not pole on, are  relativistically beamed away from the observer direction and kilonovae are intrinsically weak.  These circumstances make electromagnetic detections very difficult if the sources lie at more than $\sim$100 Mpc, as proven during the third and latest observing run (Apr 2019 - Mar 2020) of the gravitational interferometers network.   In this observing period,  two merger events possibly involving  neutron stars were reported by  the LIGO-Virgo consortium:   GW190425, caused by the coalescence of  two compact objects of masses each in the range 1.12--2.52 M$_\odot$, at $\sim$160 Mpc \citep{Abbottapj2020a},  and 
GW190814, caused by a 23 M$_\odot$ black hole merging with a compact object of 2.6  M$_\odot$ at  $\sim$240 Mpc \citep{Abbottapj2020b}.    In neither case did the search for an optical or infrared counterpart return a  positive result
\citep{Coughlin2019,Gomez2019,Ackley2020,Andreoni2020,Antier2020,Kasliwal2020},   owing presumably to the large distance and sky error areas,  although  a short GRB may have been detected by the {\it INTEGRAL} SPI-ACS  simultaneously with GW190425 \citep{Pozanenko2019}.  Note  that all coalescing stars may have been black holes, as the neutron star nature of the binary members lighter than 3  M$_\odot$ could not be confirmed.

The search for electromagnetic counterparts of gravitational  radiation signals is currently  thwarted primarily by the large  uncertainty of their localization in the sky, which is usually no more accurate than several dozens of square degrees.    Much smaller error boxes are expected to be available  when the KAGRA   (which had already joined LIGO-Virgo in the last months of the 2019-2020 observing run) and the INDIGO interferometers will operate at full regime as part of the network during the next observing run  \citep{Abbott2018}.   Observing modes, strategies, and simulations  are being  implemented  to optimize the electromagnetic multi-wavelength search and follow-up \citep{Bartos2016,Patricelli2018,Cowperthwaite2019,Graham2019,Artale2020}, and  new dedicated space-based facilities are designed with critical capabilities of large sky area coverage and rapid turnaround (e.g.  {\it ULTRASAT}, Sagiv et al., 2014; {\it THESEUS}, Amati et al., 2018, Stratta et al., 2018; {\it DORADO}, Cenko et al. 2019), to maximize the chance of detection of dim, fast-declining transients.

Finally, the possible detection of  elusive MeV and $>$GeV neutrinos associated with the kilonova \citep{KyutokuKashiyama2018} and with the GRB  \citep{Bartos2019,Aartsen2020}, respectively,  will  bring an extra carrier of information  into play, and thus complete the multi-messenger picture associated with the binary neutron star merger phenomenon.   
Gravitational waves from binary neutron star inspirals and mergers;   gamma-ray photons  -- downscattered to UV/optical/infrared light -- from  radioactive decay of  unstable nuclides of heavy elements, freshly formed after the merger;  multi-wavelength photons from non-thermal mechanisms in the relativistic jet powered by the merger remnant;  thermal and high-energy neutrinos accompanying the remnant cooling and hadronic processes in the jet, respectively,  all  collectively underpin the  role of the four  physical  interactions.  This fundamental role of compact star merger phenomenology 
thus points to the formidable opportunity offered by a multi-messenger approach: bringing the communities of   astrophysicists and nuclear physicists closer will foster that  cross-fertilization  and  inter-disciplinary  approach  that is not only beneficial, but also essential for progress in this field.

%


\section*{Acknowledgments}
The author is indebted to  T. Belloni, S. Cristallo, Th. Janka, P. Mazzali, A. Possenti, M. Tanaka, and F.  Thielemann for discussion,  and to the reviewers, F. Burgio and T. Kajino, for their critical comments and suggestions.   She acknowledges hospitality from Liverpool John Moores University;  Weizmann Institute of Science,  Rehovot, and the Hebrew University of Jerusalem, Israel;  National Astronomical Observatory of Japan,  Tokyo; Beihang University, Beijing, and Yunnan National Astronomical Observatory, Kunming, China;  Max-Planck Institute for Astrophysics and Munich Institute for Astro- and Particle Physics, Garching, Germany,  where part of this work was accomplished.

\section*{Data Availability Statement}
\begin{figure}[h!]
\begin{center}
\includegraphics[width=20cm]{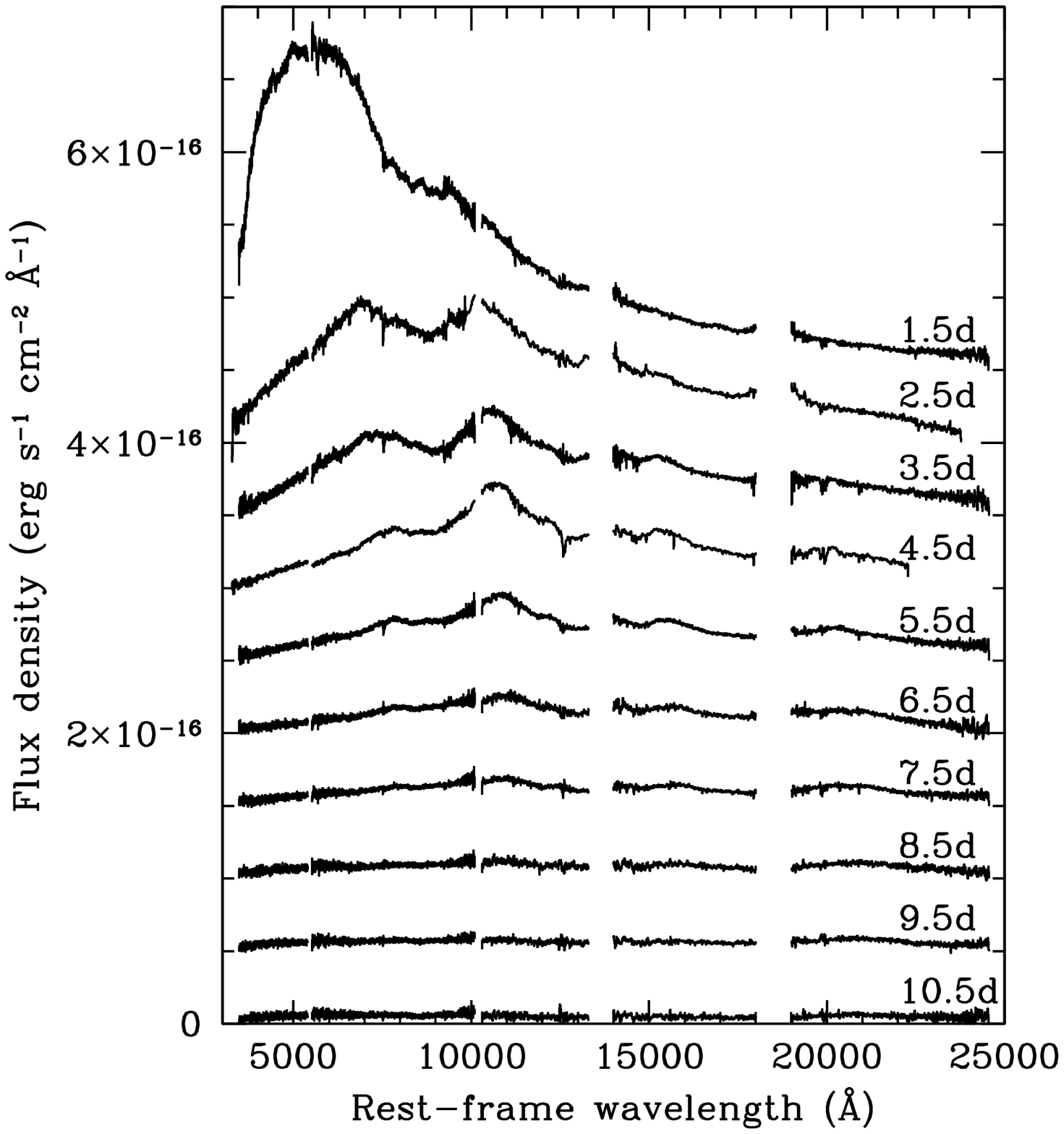}
\end{center}
\caption{ESO VLT X-Shooter spectra of the counterpart of GW170817   from   Pian et al. (2017) and Smartt et al. (2017),
at phases  indicated in days after merger time, corrected for Galactic extinction E(B-V) = 0.1 mag, de-redshifted, and offset in flux  by multiples of a  $5 \times 10^{-17}$ erg s$^{-1}$ cm$^{-2}$ \AA$^{-1}$ additive constant with respect to the 10.5d  spectrum.  Wavelength ranges of poor atmospheric transmission were blanked out.   The spectra of 19 (2.5 days) and 21 (4.5 days)  August 2017} have been re-calibrated with respect to the originally published version, courtesy of J. Gillanders, J. Selsing and S. Smartt.
\label{fig:1}
\end{figure}

The ESO VLT  X-Shooter spectra reported in Figure \ref{fig:1}, first published  in Pian et al. (2017) and Smartt et al. (2017), are available in the Weizmann Interactive Supernova Data Repository  (https://wiserep.weizmann.ac.il; Yaron \& Gal-Yam 2012).



\end{document}